\documentclass[aps,prb,letterpaper,amsmath,amssymb,superscriptaddress,reprint,
]{revtex4-1}
\pdfoutput=1 
\usepackage{graphicx}

\begin{document}
\title{Integration of antiferromagnetic Heusler compound Ru$_2$MnGe into spintronic devices}

\author{Jan Balluff}
\email{balluff@physik.uni-bielefeld.de}
\affiliation{Center for Spinelectronic Materials and Devices, Bielefeld University, D-33501 Bielefeld, Germany}
\author{Teodor Huminiuc}
\email{th881@york.ac.uk}
\affiliation{Department of Physics, University of York, York YO10 5DD, England}
\author{Markus Meinert}
\email{meinert@physik.uni-bielefeld.de}
\affiliation{Center for Spinelectronic Materials and Devices, Bielefeld University, D-33501 Bielefeld, Germany}
\email{meinert@physik.uni-bielefeld.de}
\author{Atsufumi Hirohata}
\affiliation{Department of Electronics, University of York, York YO10 5DD, England}
\author{G\"unter Reiss}
\affiliation{Center for Spinelectronic Materials and Devices, Bielefeld University, D-33501 Bielefeld, Germany}
\date{\today}

\begin{abstract}
We report on the first integration of an antiferromagnetic Heusler compound acting as a pinning layer into magnetic tunneling junctions. The antiferromagnet Ru$_2$MnGe is used to pin the magnetization direction of a ferromagnetic Fe layer in MgO based thin film tunnelling magnetoresistance stacks. The samples were prepared using magnetron co-sputtering. We investigate the structural properties by X-ray diffraction and reflection, as well as atomic force and high-resolution transmission electron microscopy. We find an excellent crystal growth quality with low interface roughnesses of 1-3\,\AA{}, which is crucial for the preparation of working tunnelling barriers. Using Fe as a ferromagnetic electrode material we prepared magnetic tunneling junctions and measured the magnetoresistance. We find a sizeable maximum magnetoresistance value of 135\%, which is comparable to other common Fe based MTJ systems.
\end{abstract}

\maketitle

\section{Introduction}
Antiferromagnets are widely used in spintronics to create a magnetically fixed ferromagnetic reference layer using the exchange bias effect \cite{Meiklejohn1956, Nogues1999}. The exchange bias effect causes a broadening and a shift of the ferromagnetic layer's hysteresis loop in the field direction. In combination with an unpinned ferromagnetic layer, magnetoresistive devices like the magnetic tunneling junction (MTJ) are designed. In addition, recently pioneering work on antiferromagnetic spintronics \cite{Jungwirth2016} was published, where antiferromagnets are used as an active component in spintronic devices. By exploiting specific symmetry properties of a material a current induced switching of its magnetic state is possible \cite{Wadley2016}. Exclusively using an antiferromagnetic material as an active component brings in the advantage of insensitivity to external magnetic fields e.g. for data storage. Thus, antiferromagnets play an important role in the field of spintronics. Especially the widely used antiferromagnetic IrMn or PtMn are, however, costly and rare. In conjunction with the rising field of antiferromagnetic spintronics suitable, novel antiferromagnetic materials are of increasing interest.

Heusler compounds are a ternary material class of the type X$_2$YZ, where the basic crystal structure is a four-atom basis in an fcc lattice (space group Fm$\overline{\mathrm{3}}$m, prototype Cu$_2$MnAl). They are very versatile rendering them interesting for a wide range of applications \cite{Graf2011}. Ferro- and ferrimagnetic Heusler compounds are extensively studied \cite{Bai2012} as they provide large magnetoresistance ratios \cite{Liu2012} in giant or tunnelling magnetoresistance (GMR\cite{Baibich1988, Binasch1989} and TMR\cite{Gregg2002, Chappert2007}, respectively) devices. However, to the best of our knowledge no work on spintronic devices using an antiferromagnetic Heusler compound as a pinning layer has been reported so far. Due to the matching crystal structure a combination of antiferromagnetic and ferromagnetic Heusler compounds can lead to high quality TMR stacks.

We study the integration of the recently investigated \cite{Balluff2015, Fukatani2013} antiferromagnetic Heusler compound Ru$_2$MnGe (RMG) into MTJ spin valves. Within our previous work we have already shown that a sizeable exchange bias effect of up to 600\,Oe is found in RMG / Fe bilayers \cite{Balluff2015}. Furthermore, we measured the blocking temperature, at which the exchange bias vanishes, to be $T_\mathrm{B}=130$\,K. Within this work, we prepared RMG based thin film devices using dc and rf magnetron co-sputtering as well as electron beam evaporation. We compare measurements of the thin film roughness and crystal growth quality by using methods of X-ray diffraction (XRD), atomic force microscopy (AFM) and high resolution transmission electron microscopy (HR-TEM). Furthermore, the resulting TMR amplitudes of our devices as a function of different annealing temperatures are investigated to improve effect sizes and especially examine the applicability by an investigation of the tunnelling barrier quality.

\section{Experimental details}
\begin{figure}[t]
\includegraphics[width=8.2cm]{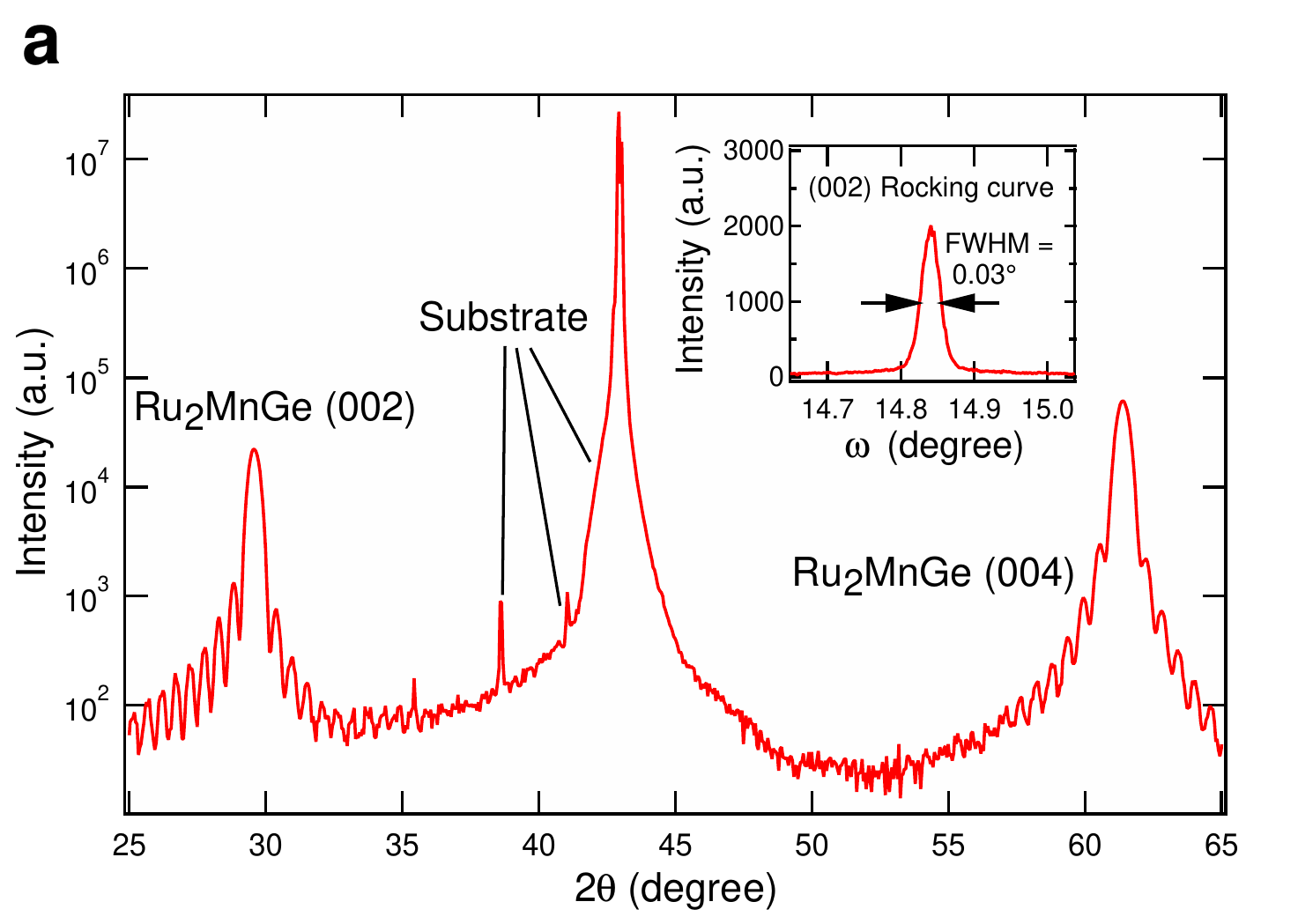}
\includegraphics[width=8.2cm]{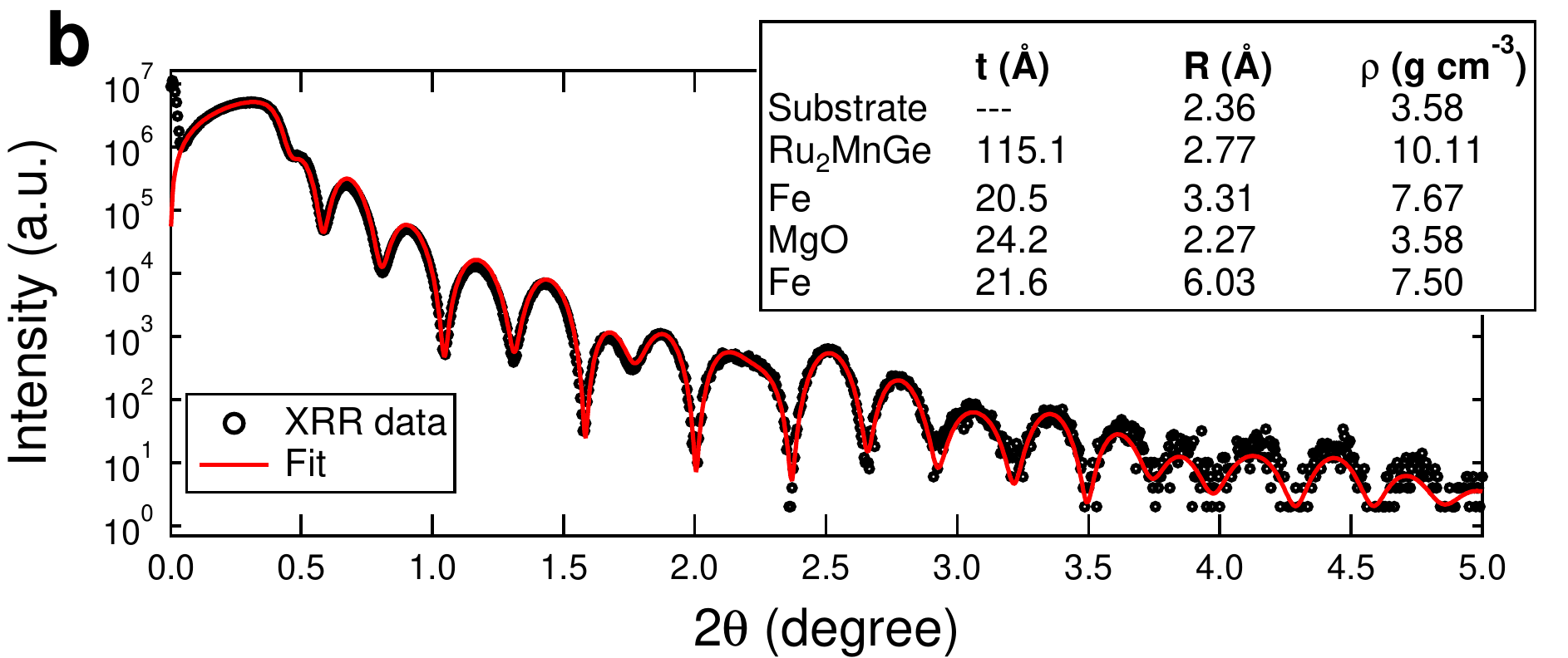}
\caption{X-ray diffraction analysis. \textbf{(a)}: High resolution diffraction pattern of a single 20\,nm RMG layer. Laue oscillations at both (002) and (004) indicate excellent epitaxial crystal growth supported by the narrow rocking curve shown in the inset. \textbf{(b)}: X-ray reflectivity of a full TMR stack. The black dots are the measured data whereas the red solid line is a fit according to the Parratt formalism. Parameters obtained by the fit are given in the inset table.}
\label{fig:xrd}
\end{figure}
Our RMG layers were prepared using magnetron co-sputtering from elemental targets, where the Ar working pressure is typically $2.3\cdot10^{-3}$\,mbar during the process. The base pressure of the sputter deposition system is better than $10^{-8}$\,mbar. Adjusting the magnetron power allows precise control of the stoichiometry, which was checked using X-ray fluorescence and is typically accurate within $<$1\%at. The RMG layer was sputter-deposited on MgO single crystalline substrates with the epitaxial relation RMG[100] $\parallel$ MgO[110]. The lattice mismatch is 0.5\%, so no buffer layer was used. The layer was deposited at a substrate temperature of 500$^\circ$C. After deposition, the sample was further annealed in-situ at the same temperature for one hour and then cooled down to ambient temperature. A TMR stack in the form of Fe 2\,nm / MgO 2\,nm / Fe 2\,nm was deposited at room temperature. All layers were deposited by magnetron sputtering except the MgO tunneling barrier, which was deposited using an electron beam evaporator with a deposition rate of approximately $0.1\,$\AA{}/s. As an electrical contact, a layer of Ta 3\,nm / Ru 5\,nm was deposited on top of the TMR stack.

In a first step, the samples were analyzed by X-ray reflectivity and diffraction in a Philips X'Pert Pro MPD diffractometer with Bragg-Brentano optics operated with Cu K$_\alpha$ radiation. Further characterization of the samples regarding interface roughness was done using X-ray reflectivity (XRR) and AFM. XRR measurements were done up to $2\theta = 5^\circ$ and fitted according to the Parratt formalism \cite{Parratt1954}. AFM images were recorded using a Bruker Multimode 5 microscope operated in tapping mode. Magnetic analysis of the exchange bias provided by RMG is found elsewhere \cite{Balluff2015}.

The tunneling barrier was investigated by cross-sectional HR-TEM using a JEOL JEM-2200FS electron microscope operating at 200kV and equipped with a CEOS image aberration corrector. The samples were prepared by cutting and manually grinding the samples before further processing. The thinned samples were Ar ion milled to electron transparency with a Gatan Precision Ion Polishing System using a temperature controlled stage in order to prevent intermixing at the interfaces.

For the final investigation of MTJ devices the samples were patterned in a standard UV lithography process in combination with secondary ion mass spectroscopy controlled Ar ion beam etching. Square nano pillars of $7.5\times7.5\,\mathrm{\mu}$m$^2$ were prepared. The RMG layer is used as a bottom contact for all MTJ cells. Samples were mounted on a chip carrier for electrical measurements and contacted by Au bonding wire using ball and wedge bonding. The magnetoresistance of the TMR devices was measured in a closed-cycle He cryostat.


\section{Results}
\begin{figure}[t]
\includegraphics[width=4.0cm]{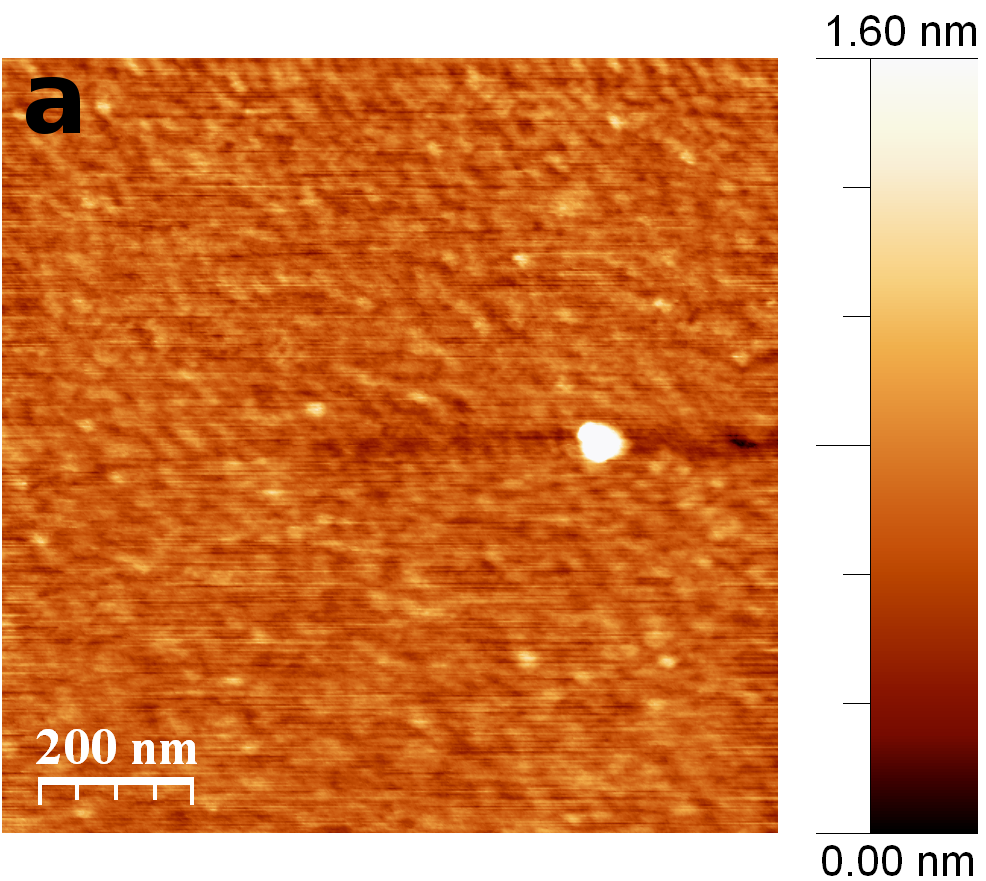}
\hspace{0.2cm}
\includegraphics[width=4.0cm]{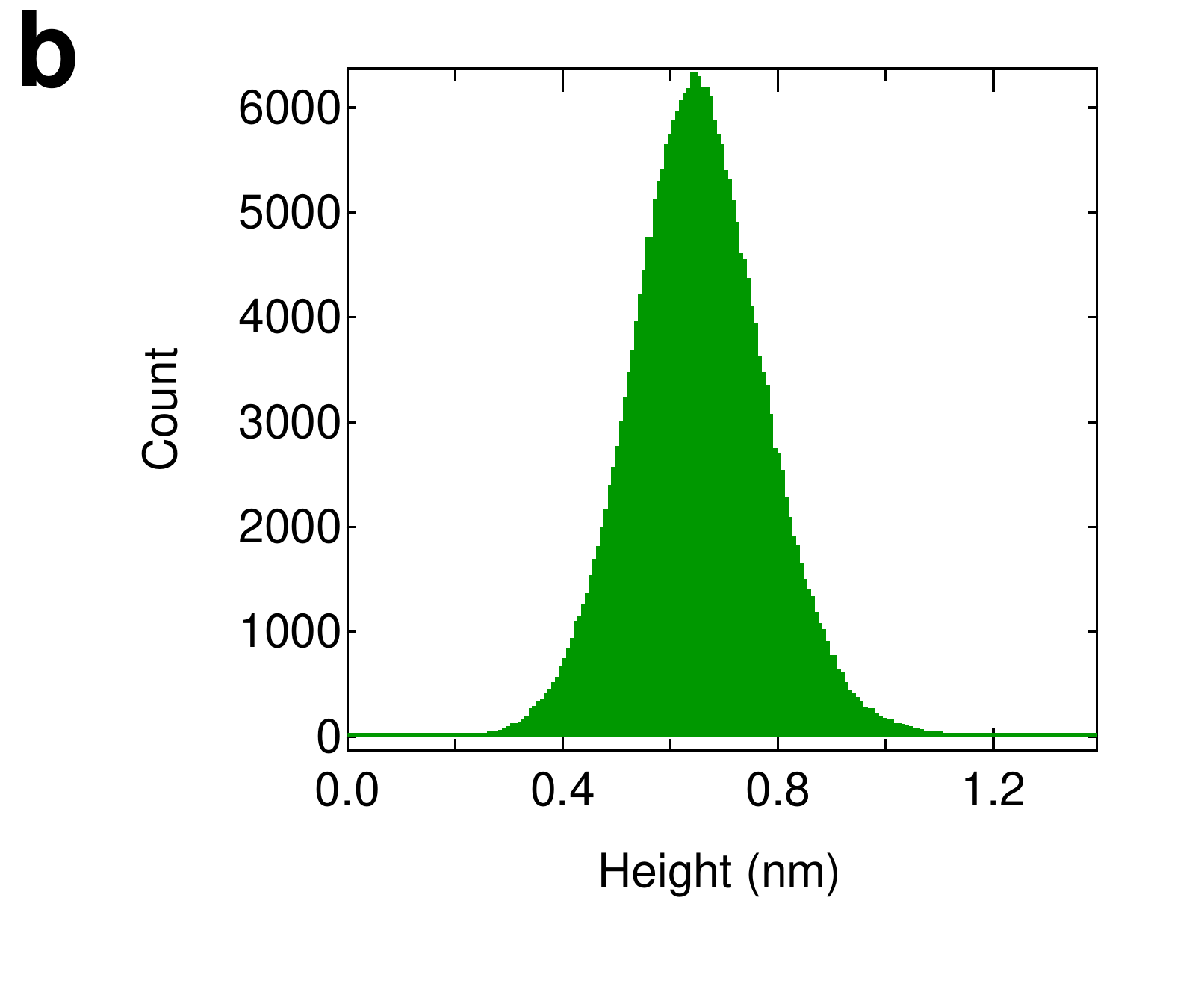}\\
\includegraphics[width=4.0cm]{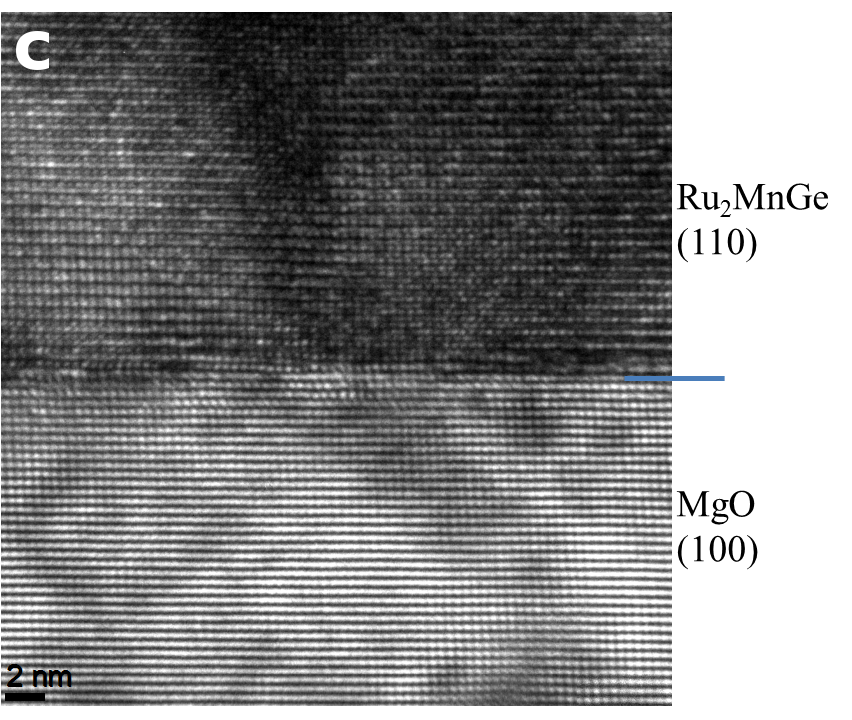}
\hspace{0.2cm}
\includegraphics[width=4.0cm]{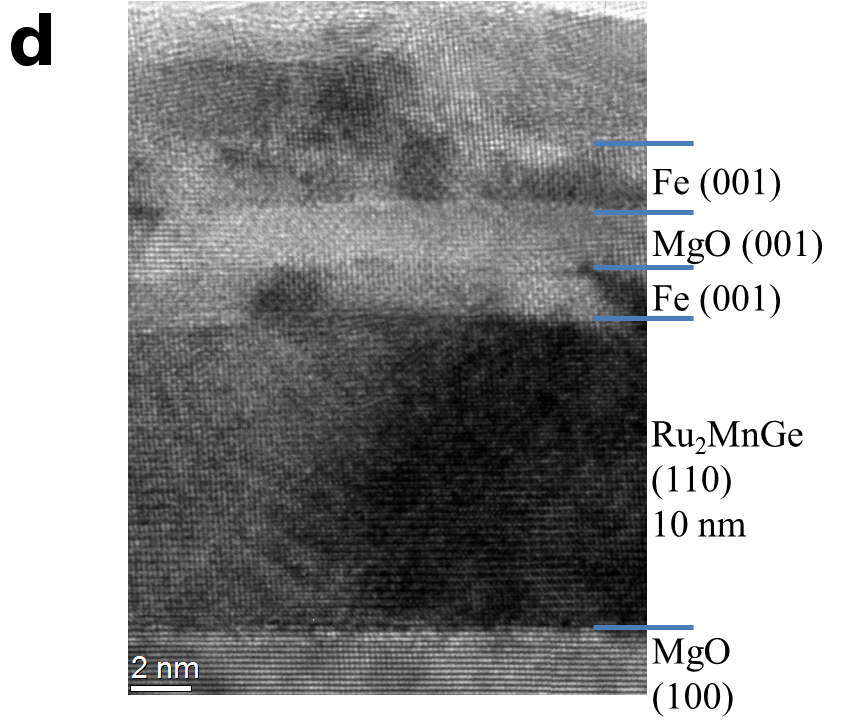}
\caption{AFM and HR-TEM images of an MTJ. \textbf{(a)}: AFM image of an MTJ with Fe electrondes. The white dot is due to contamination. A smooth and homogenous sample surface is found. \textbf{(b)}: Height distribution histogram of the AFM image. \textbf{(c)}: HR-TEM image of the interface between MgO and RMG showing the epitaxial crystal growth. \textbf{(d)}: HR-TEM image of a full MTJ cross section. A clean, crystalline MgO barrier is clearly visible.}
\label{fig:afm}
\end{figure}
The RMG layer shows excellent crystalline growth. The diffraction pattern for a 20\,nm thick layer without a TMR stack is shown in in Fig. \ref{fig:xrd}a. Here, the expected (002) and (004) peaks for the Heusler structure are found. Both show pronounced Laue oscillations, which are an indication for homogeneous crystal growth. This is further supported by a narrow rocking curve with a full width at half maximum (FWHM) of $<0.03^\circ$ (shown in the inset). This value is limited by the divergence of the diffractometer optics. The results obtained by XRR for a RMG / Fe / MgO / Fe TMR stack are plotted in Fig. \ref{fig:xrd}b. Here, the measured data as black dots is shown in conjunction with a fit in red. The fit precisely matches the measured data even up to large angles. The resulting layer thickness as well as roughness and density of the RMG, Fe and MgO layers are given in the graph as well. 

As indicated by the XRR and XRD analysis, high quality crystal growth is obtained without the necessity of external sample treatment such as further post annealing. The final fit parameter values as given in Fig. \ref{fig:xrd}b indicate a very low roughness of 2-3\,\AA{} for the interfaces. For the upper Fe layer, a slightly increased thickness and lower density is found, which is attributed to the increased roughness of 6\,\AA{}.

\begin{figure}[t]
\includegraphics[width=7.8cm]{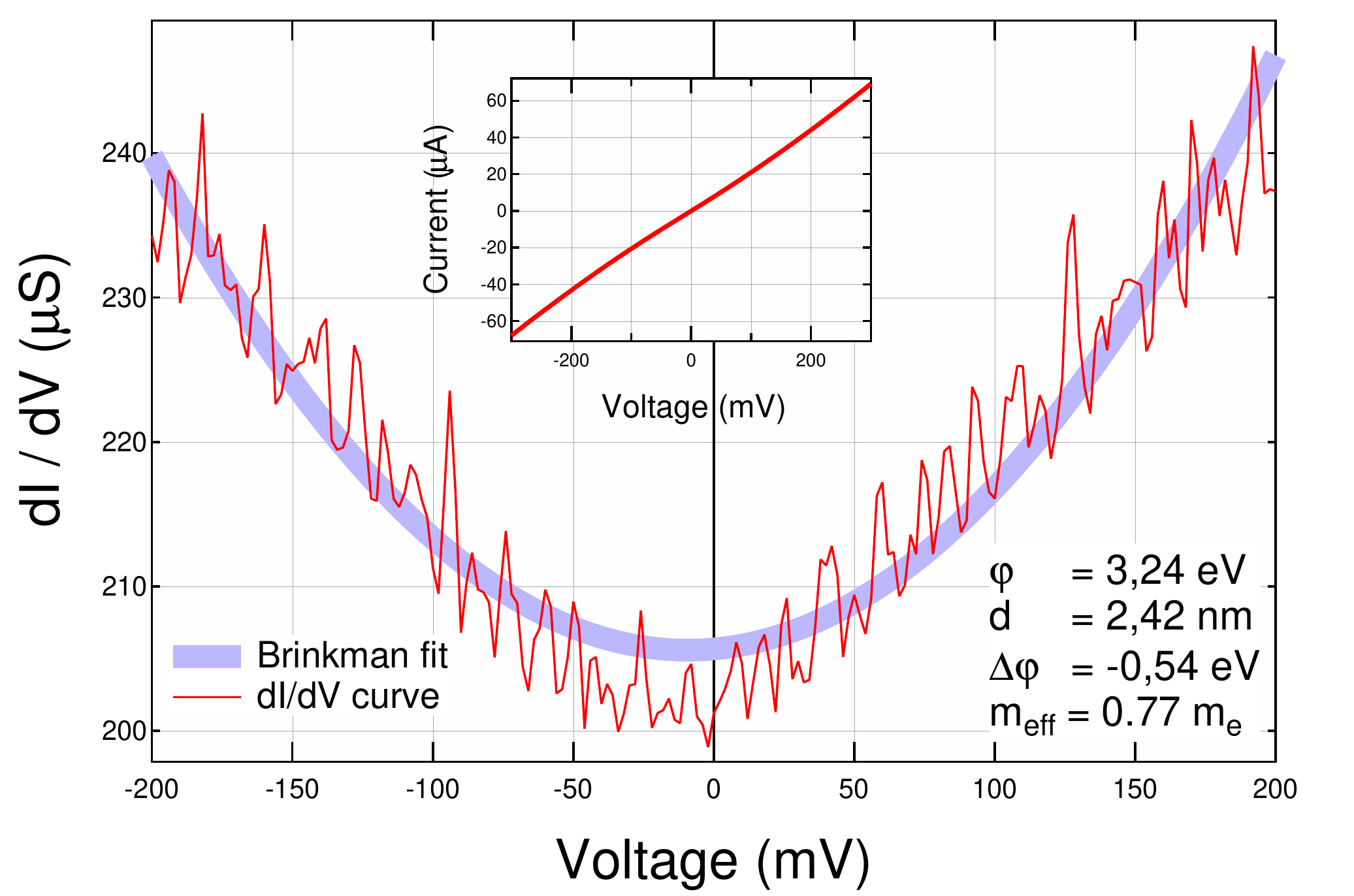}
\caption{I-V measurement (inset) and its numerical derivative $dI/dV$ (thin line) of a TMR stack with a Brinkman fit (thick, shaded line). The final fit parameters barrier height $\varphi$, barrier thickness $d$, barrier asymmetry $\Delta\varphi$ and effective electron mass $m_\mathrm{eff}$ are given.}
\label{fig:didv}
\end{figure}
An AFM image of a full TMR stack's surface is shown in Fig. \ref{fig:afm}a. The image shows a smooth sample surface without cluster or island nucleation. The white dot is due to contamination and not attributed to the sample. In Fig. \ref{fig:afm}b the height distribution across the AFM image is given. The low roughness obtained from the XRR measurements is confirmed by this measurement. Here, we find a RMS roughness of 1.3\,\AA{} (the contamination is excluded from this calculation).

Fig. \ref{fig:afm}c and d show HR-TEM cross section images of the sample. The epitaxial growth of the antiferromagnetic RMG is confirmed via the sharp substrate/Heusler alloy interface as seen in Fig. \ref{fig:afm}c with no defects observed in the bulk of the material. This agrees with the crystallographic studies done by XRD. In the RMG layer the ordered Heusler structure is visible by the alternating planes of Ru and Mn-Ge. The $1:1/\sqrt{2}$ relation of the unit cell dimensions are as expected for the RMG [110] interface. Fig. \ref{fig:afm}d shows all layers with atomic smooth growth throughout the whole TMR stack. The MgO tunnel barrier and the two ferromagnetic layers show very good crystalline quality throughout and lattice matched deposition at the bcc Fe (001)/MgO (001)/Fe (001) tunnelling interface. The visible 11-12 atomic layers of MgO correspond to a barrier thickness of $23.2-25.3$\,\AA{} ($a_\mathrm{MgO} = 4.21$\,\AA{}) confirming the results obtained by XRR. The slight increase in roughness at the interface between the top Fe layer and capping layer is confirmed as observed in the XRR measurements. This does, however, not affect the quality of the MgO barrier.

We investigated the tunneling magnetoresistance of square nano pillar MTJs. Measuring the I-V characteristic as a function of V at room temperature reveals a working tunneling barrier. Applying a Brinkman fit \cite{Brinkman1970} to the numerical derivative $dI/dV$ allows to determine tunneling barrier height $\varphi$, asymmetry $\Delta\varphi$ and thickness $d$. The inset in Fig. \ref{fig:didv} shows the experimental I-V data. The numerically evaluated $dI/dV$ curve (thin line) is shown in the main plot of Fig. \ref{fig:didv} as well as the Brinkman-fit (thick, shaded line). The effective electron mass $m_\mathrm{eff}$ is a free parameter in this model. As we know the barrier thickness exactly from XRR and HR-TEM, we adjust $m_\mathrm{eff}$ to obtain the correct value. The final fit parameters given in Fig. \ref{fig:didv} are reasonable considering the MgO band gap of 7.8\,eV \cite{Taurian1985}.

\begin{figure}[t]
\includegraphics[width=4.1cm]{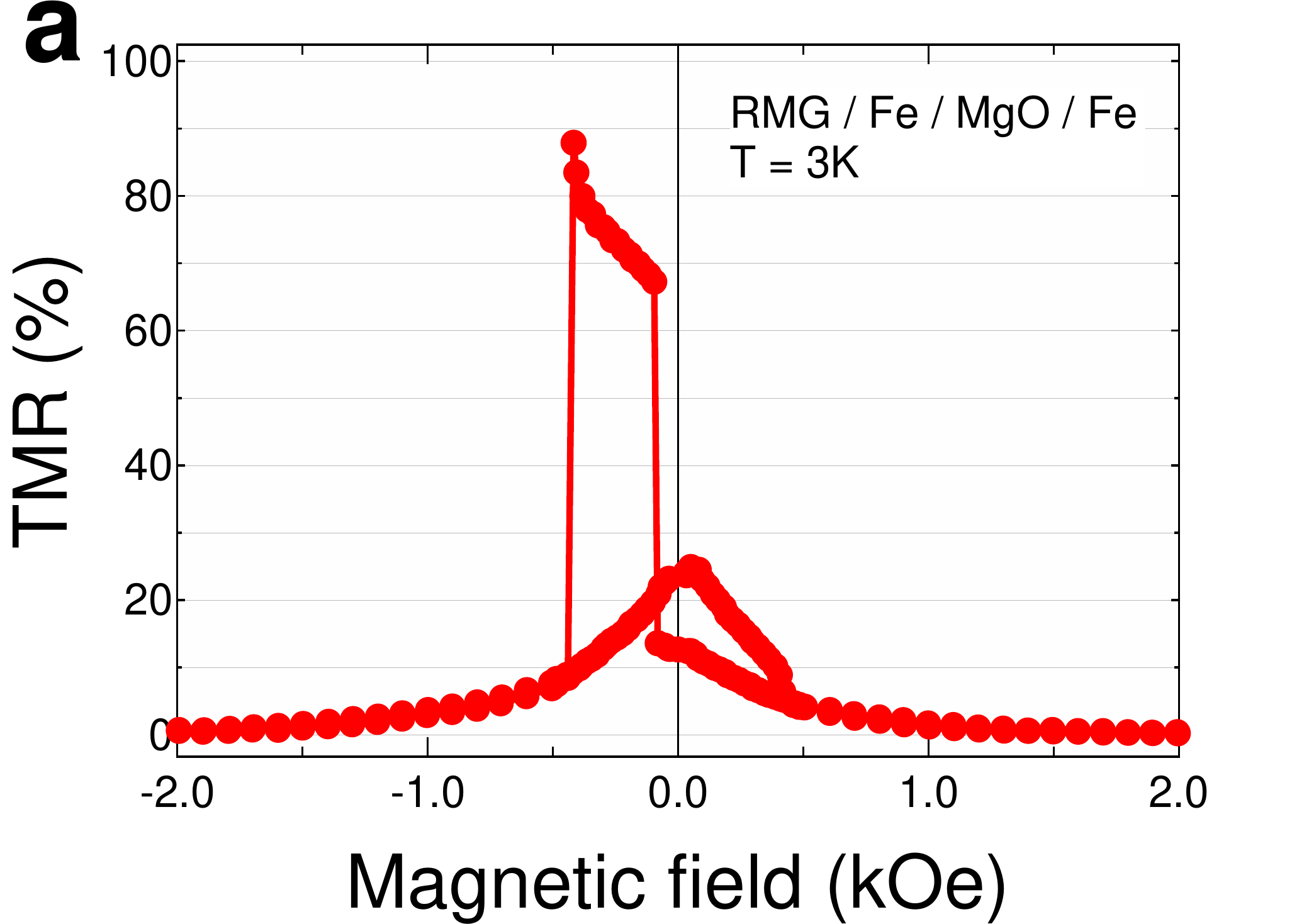}
\includegraphics[width=4.1cm]{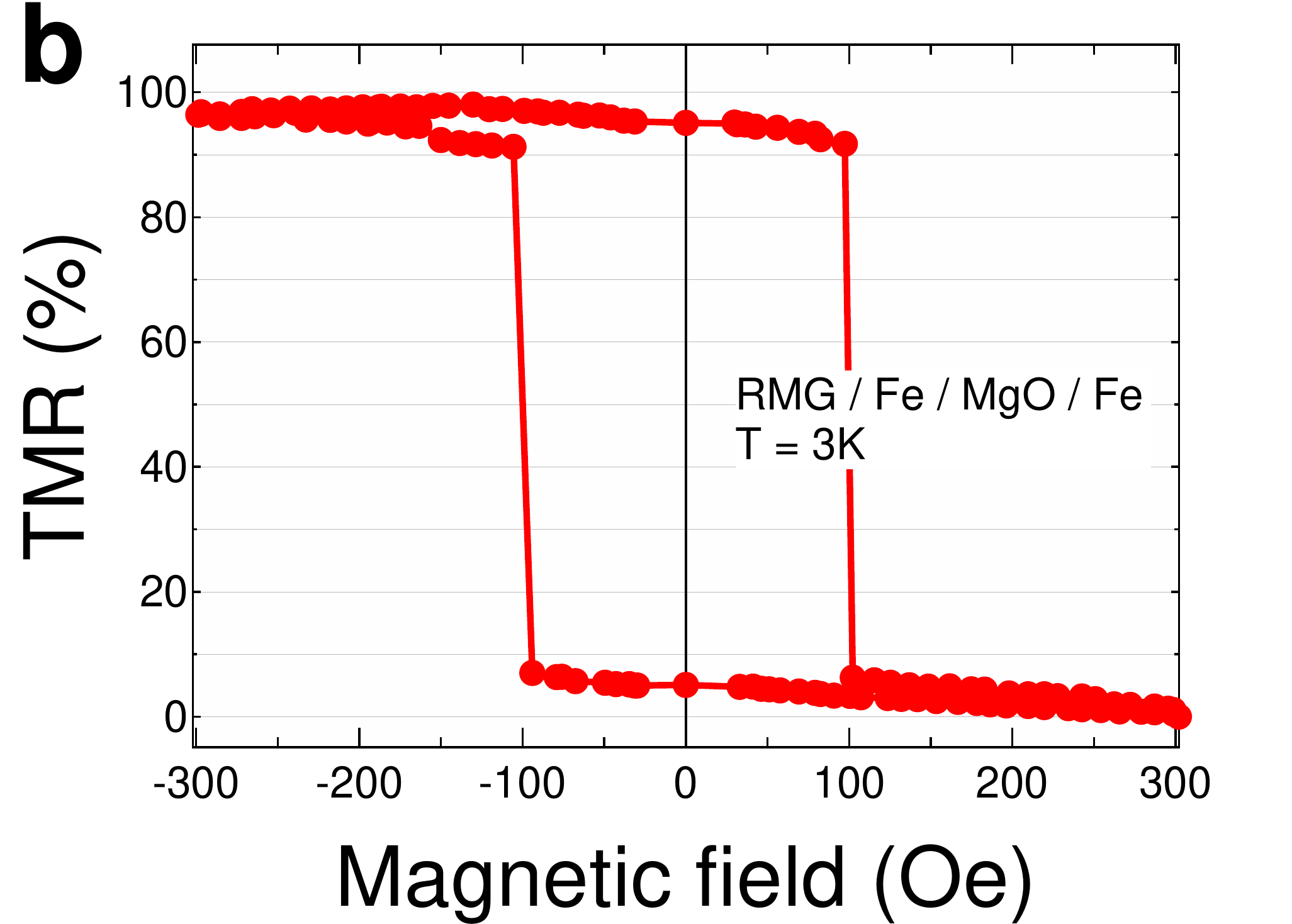}
\caption{Tunneling magnetoresistance of a TMR stack recorded at 3\,K. \textbf{(a)}: Major loop switching the whole stack. The coercive fields of the two ferromagnetic layers are similar in the positive field regime, hence no sharp switching is observed. \textbf{(b)}: Minor loop only switching the unpinnend ferromagnetic layer. A sharp, square swichting with an amplitude of about 100\% is observed.}
\label{fig:loops}
\end{figure}

Due to the low blocking temperature $T_\mathrm{B}=130\,K$ of the RMG / Fe bilayer system, the samples are cooled down in a closed-cycle He cryostat for magnetoresistive characterization. During the cooldown, a magnetic field of 4\,T was applied. After cooling down, the magnetoresistance is measured by applying a constant voltage of $U=10$\,mV across the MTJ and sweeping the magnetic field parallel to the sample. The corresponding loops are shown in Fig. \ref{fig:loops} where the magnetoresistance $\mathrm{TMR} = (R_\mathrm{ap} - R_\mathrm{p})/R_\mathrm{p}$ is plotted against the external magnetic field. $R_\mathrm{ap}$ and $R_\mathrm{p}$ are the resistance values in antiparallel (ap) and parallel (p) states, respectively. In the asymmetric major loop (Fig. \ref{fig:loops}a), the shifted hysteresis of the exchange biased Fe layer is clearly seen, leading to a distinct switching of the two Fe electrodes. The exchange bias observed in the full structured TMR stacks is reduced by a factor of 2-3 to about 250\,Oe compared to the previously investigated RMG / Fe bilayers \cite{Balluff2015}. The quality of the switching is limited due to the UV lithography process and the corresponding large size of the nano pillars. The minor loop shown in Fig. \ref{fig:loops}b, however, shows a nearly perfect square switching. The TMR has a sizeable value of about 100\%.

\begin{figure}[t]
\includegraphics[width=4.1cm]{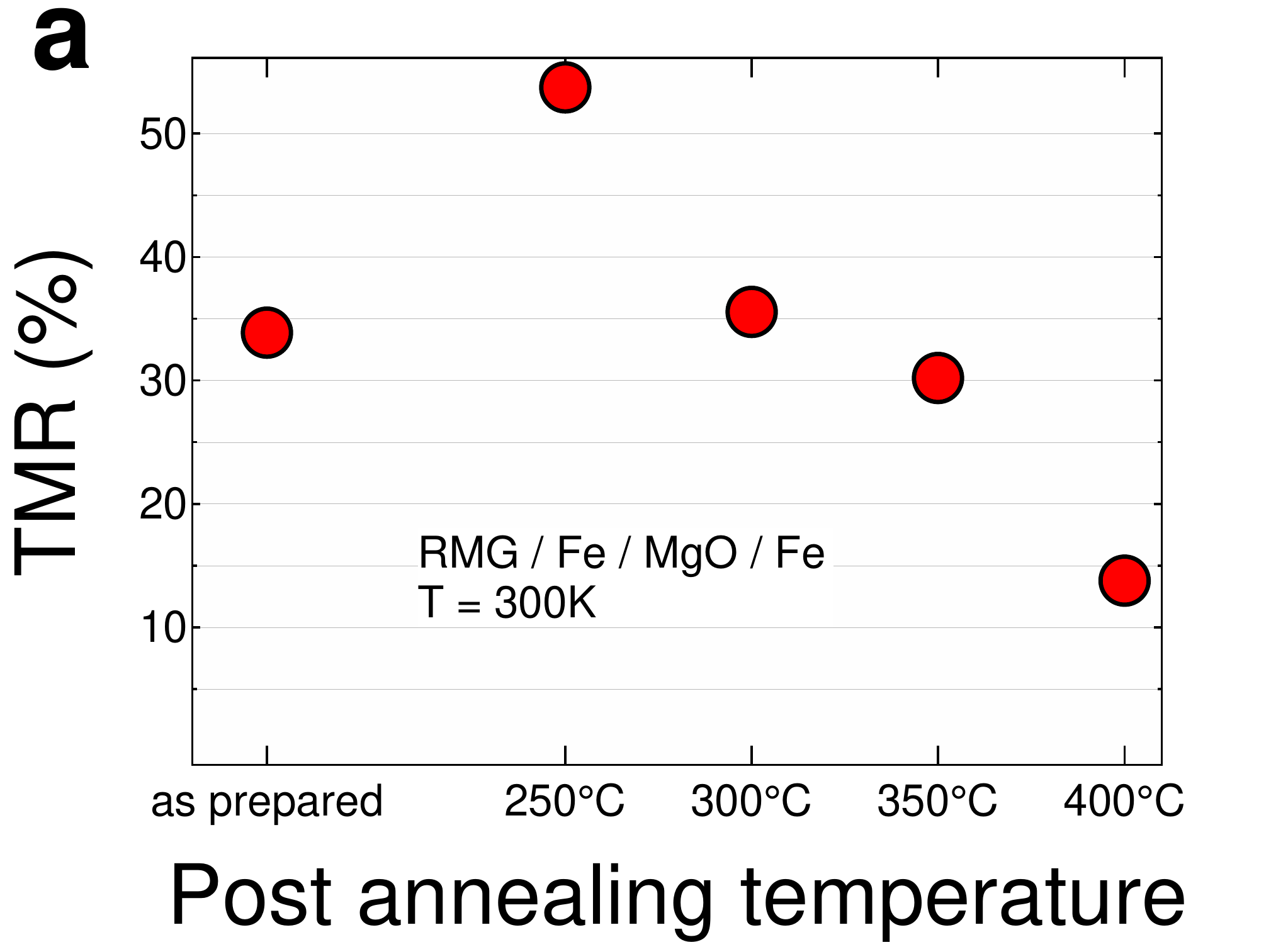}
\includegraphics[width=4.1cm]{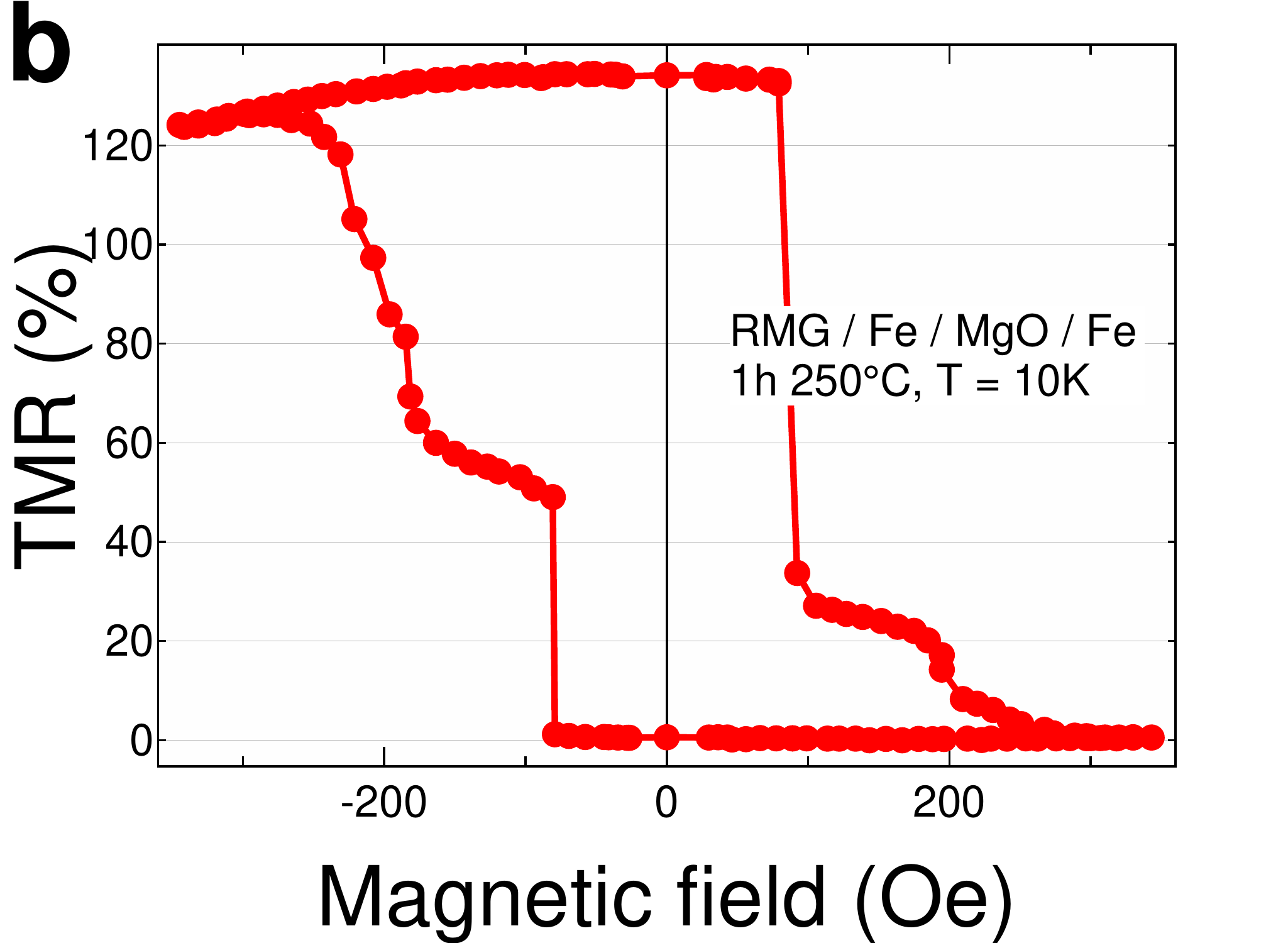}
\caption{Effects of ex-situ post anneal of full TMR stacks. \textbf{(a)}: TMR amplitudes for the as-prepared sample as well as different annealing temperatures for 60\,min recorded at 300\,K. \textbf{(b)}: Minor loop recorded at 10\,K for a sample annealed at 250$^\circ$C. The TMR amplitude is enhanced to about 135\%, but the loop shows multi domain switching.}
\label{fig:tmr_pa}
\end{figure}
We further investigated the TMR after ex-situ post annealing samples in a vacuum furnace at $10^{-7}$\,mbar prior to lithography. The samples were annealed at 250$^\circ$C to 400$^\circ$C in steps of 50$^\circ$C, which are typical post annealing temperatures for TMR spin valves. Samples for post annealing are prepared with a slightly increased thickness (3\,nm) of the top Fe electrode. Due to the asymmetry of the two ferromagnetic layers, a comparable measurement of the TMR in the unpinned state at room temperature is possible. Low temperature measurements confirmed that this does not affect the TMR effect size. The TMR values measured at room temperature are compared to the as-prepared sample. The results are shown in Fig. \ref{fig:tmr_pa}a.  The highest TMR value is observed for post annealing at 250$^\circ$C, whereas for 300$^\circ$C we found a TMR value comparable to the as-prepared sample. Any further increase of the annealing temperature led to smaller effect sizes. Thus, we investigated a sample annealed at 250$^\circ$C at low temperatures. The exchange bias compared to the as-prepared sample is increased to 380\,Oe. A minor loop recorded for this sample is shown in Fig. \ref{fig:tmr_pa}b. We observe a clear enhancement in the TMR amplitude to 135\% compared to the unannealed sample. However, multidomain switching is clearly visible in the graph, which is unfavourable for a clean switching of the spin valve. This is induced by the post annealing of the whole layered stack. We explain this by further crystallization effects affecting the grain sizes of the upper Fe electrode, also supported by its increased roughness, as well as Mn diffusion from the RMG layer into the TMR stack.

\section{Conclusion}
We have successfully demonstrated the first integration of an antiferromagnetic Heusler compound as a pinning layer into magnetic tunneling junctions. Investigation of the sputtered thin film multilayers RMG / Fe MgO / Fe by X-ray techniques revealed an excellent crystalline growth combined with a low roughness. Especially, smooth surfaces can be obtained directly in the sputtering process without the necessity of ex-situ treatment, which is confirmed by AFM measurements. A more detailed insight of the MTJ quality is given by HR-TEM investigations. Here, we find the epitaxial growth of the RMG layer on the MgO substrate without any defects. Also, a good quality tunneling barrier throughout the crystal is found, not affected by interface roughness. Our investigations of the magnetoresistance at low temperatures revealed working MTJ nano pillars with a sharp, square-shaped switching in the minor loop of 100\,\% signal amplitude. The quality of the switching in the major loop is still subject to improvements and mainly limited to the UV lithography process, which limits the device size. We found a decent increase in signal amplitude to 135\,\% as well as in exchange bias when annealing samples at 250$^\circ$C. The effect amplitudes we obtained in the RMG-based TMR system are comparable to similar Fe / MgO / Fe systems \cite{Faure2003}. An ex-situ treatment can improve the TMR effect size. Further investigations will include different electrode materials, which may behave differently under post annealing conditions. Especially, our investigation can establish a basis for "all-Heusler" MTJs with MgO tunneling barriers. Due to the matching crystal structure and giant effect sizes already found in MTJs using Heusler compounds as an electrode material \cite{Liu2012}, this is an appealing future task. All in all, the antiferromagnetic RMG Heusler compound is a promising material due to the ease of fabrication. The compound itself or similar related Heusler compounds may be useful in future applications, or even in the new field of antiferromagnetic spintronics.

\section{Acknowledgement}
The research leading to these results has received funding from the European Union Seventh Framework Programme (FP7/2007-2013) under grant agreement no. NMP3-SL-2013-604398.

\end{document}